\documentclass[superscriptaddress,twocolumn,showpacs,preprintnumbers,amsmath,amssymb]{revtex4}

\usepackage{graphicx}
\usepackage{dcolumn}
\usepackage{bm}

\begin{document}

\preprint{}

\title{Hole Binding around Ni Impurity in Cuprates}

\author{Kenji Tsutsui}
\affiliation{Synchrotron Radiation Research Center, Japan Atomic Energy Agency, Hyogo 679-5148, Japan.}
\author{Atsushi Toyama}%
\affiliation{Institute for Materials Research, Tohoku University, Sendai 980-8577, Japan.}
\author{Takami Tohyama}%
\affiliation{Yukawa Institute for Theoretical Physics, Kyoto University, Kyoto 606-8502, Japan.}
\author{Sadamichi Maekawa}%
\affiliation{Institute for Materials Research, Tohoku University, Sendai 980-8577, Japan.}
\affiliation{CREST, Japan Science and Technology Agency (JST), Tokyo 102-0075, Japan.}

\date{\today}

\begin{abstract}
We examine the influence of Ni impurity in cuprates on the distribution of hole carriers by performing numerically exact diagonalization calculations for a model consisting of Cu$3d$, Ni$3d$, and O$2p$ orbitals.
Using realistic parameters for the system, we find that a hole is predominantly bound to O$2p$ orbitals around the Ni impurity forming the Zhang-Rice doublet.
This imposes strong restrictions on modeling Ni-substituted cuprates.
We propose a resonant inelastic x-ray scattering experiment for Ni $K$-edge to confirm hole binding around the Ni impurity.
\end{abstract}

\pacs{74.72.-h, 71.10.Fd, 71.55.-i}
\maketitle
Atomic substitutions for copper in high-temperature cuprate superconductors induce significant impacts on macroscopic and local physical properties of the cuprates~\cite{alloul}.
Divalent transition metal ions such as zinc and nickel have been frequently substituted for Cu.
It is well known that nonmagnetic Zn impurities suppress the superconducting transition temperature more strongly than magnetic Ni impurities.
Concerning magnetic properties, neutron scattering experiments have shown that Zn impurities enhance the antiferromagnetic (AF) correlation length only slightly, whereas Ni impurities suppress incommensurate peaks strongly and stabilize the Neel ordering~\cite{birgeneau}.
Such contrasting behaviors have been discussed in connection with the difference of spin state of Zn$^{2+}$ with spin $S=0$ and Ni$^{2+}$ with $S=1$~\cite{alloul}.
For Ni substitution, however, it has been argued from several experiments~\cite{bhat,nakano,hiraka2,matsuda,hiraka3} that a Ni impurity attracts a hole, giving Ni$^{3+}$ with $S=\frac{1}{2}$.
More recently, it has been suggested from x-ray-absorption-fine-structure (XAFS) measurements that the Ni impurity forms Ni$^{2+}\underline{L}$ state with $S=\frac{1}{2}$ ($\underline{L}$ represents a ligand hole), i.e., Zhang-Rice (ZR) doublet state, binding a hole on neighboring oxygen orbitals ~\cite{hiraka}.
The formation of the doublet may explain several experimental facts for Ni-substituted cuprates as discussed in Ref.~\cite{hiraka}: small moment of Ni impurity~\cite{xiao,mendels}, weak suppression of coherence peaks~\cite{hudson}, reduction of magnetic resonance energy~\cite{sidis}, enhancement of pseudogap energy~\cite{pimenov}, and so on.
In spite of accumulating experimental evidences of hole binding around Ni impurity, there is no theoretical support clarifying the formation of the ZR doublet as far as we know.

In this Letter, we show that the ZR doublet is certainly formed at Ni site embedded in the CuO$_2$ plane, based on exact diagonalization calculations for small clusters with realistic parameter values.
This theoretical result together with experimental ones mentioned above puts restrictions on theoretical models of Ni-substituted cuprates.
We propose a site-selective experimental technique to confirm the presence of the ZR doublet, which is resonant inelastic x-ray scattering (RIXS) experiment for Ni $K$-edge.
We predict low-energy structures inside the Mott gap, which can be direct evidence of hole binding around Ni impurity.

For Cu and O ions, we consider Cu$3d_{x^2-y^2}$ and O$2p_\sigma$ orbitals.
An additional orbital $3d_{3z^2-r^2}$ is included on Ni ion substituted for Cu. Apical oxygen $2p_z$ orbitals below/above the Ni ion are also taken into account.
By including hoppings between $3d$ and $2p$ orbitals and the Coulomb interactions among $3d$ orbitals, the Hamiltonian in the hole representation is given by $H_{dp}=H_T+H_d$ with
\begin{eqnarray}\label{dp1}
H_T&=&T_{pd}\sum_{\mathbf{i}\sigma}d^\dag_{\mathbf{i}\sigma}
\left(p_{\mathbf{i-\frac{x}{2}}\sigma}-p_{\mathbf{i+\frac{x}{2}}\sigma}
-p_{\mathbf{i-\frac{y}{2}}\sigma}+p_{\mathbf{i+\frac{y}{2}}\sigma}\right)   \nonumber\\
&-&T'_{pd}\sum_{\mathbf{i_0}\sigma}d'^\dag_{\mathbf{i_0}\sigma}
\left(p_{\mathbf{i_0-\frac{x}{2}}\sigma}-p_{\mathbf{i_0+\frac{x}{2}}\sigma}
+p_{\mathbf{i_0-\frac{y}{2}}\sigma}-p_{\mathbf{i_0+\frac{y}{2}}\sigma}\right)   \nonumber\\
&+&\alpha T''_{pd}\sum_{\mathbf{i_0}\sigma}d'^\dag_{\mathbf{i_0}\sigma}
\left(p_{\mathbf{i_0-z}\sigma}-p_{\mathbf{i_0+z}\sigma}\right)
+\mathrm{H.c.}   \nonumber\\
&+&\varepsilon_d\sum_{\mathbf{i}\neq\mathbf{i}_0\sigma}n^d_{\mathbf{i}\sigma}
+\varepsilon_\mathrm{Ni}\sum_{\mathbf{i}_0\sigma\gamma=d,d'}n^\gamma_{\mathbf{i}_0\sigma}
+\varepsilon_p\sum_{\mathbf{i}\mathbf{\delta}\sigma}n^p_{\mathbf{i+\frac{\delta}{2}}\sigma},\label{dp}
\\
H_d&=&U_d\sum_{\mathbf{i}\neq\mathbf{i}_0}n^d_{\mathbf{i}\uparrow}n^d_{\mathbf{i}\downarrow}
+U_\mathrm{Ni}\sum_{\mathbf{i}_0\gamma}n^\gamma_{\mathbf{i}_0\uparrow}n^\gamma_{\mathbf{i}_0\downarrow}   \nonumber\\
&+&U'_\mathrm{Ni}\sum_{\mathbf{i}_0\sigma\sigma'}n^d_{\mathbf{i}_0\sigma}n^{d'}_{\mathbf{i}_0\sigma'}
+K_\mathrm{Ni}\sum_{\mathbf{i}_0\sigma\sigma'}d^\dag_{\mathbf{i}_0\sigma}d'^\dag_{\mathbf{i}_0\sigma'}d_{\mathbf{i}_0\sigma'}d'_{\mathbf{i}_0\sigma}   \nonumber\\
&+&K_\mathrm{Ni}\sum_{\mathbf{i}_0}\left(
d^\dag_{\mathbf{i}_0\uparrow}d^\dag_{\mathbf{i}_0\downarrow}d'_{\mathbf{i}_0\downarrow}d'_{\mathbf{i}_0\uparrow}+\mathrm{H.c.}\right),
\end{eqnarray}
where the summation of $\mathbf{i}$ runs over all of Cu and Ni sites, $\mathbf{i}_0$ denotes Ni sites, $\mathbf{x}$ ($\mathbf{y}$) is the vector connecting neighboring Cu ions along the $x$ ($y$) direction, $\mathbf{i_0\pm z}$ represents apical oxygen sites above/below Ni sites, and $\delta$ runs over $\mathbf{x}$, $\mathbf{y}$, and $2\mathbf{z}$.
The operator $d_{\mathbf{i}\sigma}$ ($d'_{\mathbf{i}_0\sigma}$) is the annihilation operator for a $3d_{x^2-y^2}$ ($3d_{3z^2-r^2}$) hole with spin $\sigma$ at site $\mathbf{i}$ ($\mathbf{i}_0$), $p_{\mathbf{i}\pm\frac{\delta}{2}\sigma}$ is the annihilation operator for a $2p$ hole at site $\mathbf{i}\pm\frac{\delta}{2}$ with spin $\sigma$.
$n^\gamma_{\mathbf{i}_0\sigma}=\gamma^\dag_{\mathbf{i}_0\sigma}\gamma_{\mathbf{i}_0\sigma}$ and $n^p_{\mathbf{i}+\frac{\delta}{2}\sigma}=p^\dag_{\mathbf{i}+\frac{\delta}{2}\sigma}p_{\mathbf{i}+\frac{\delta}{2}\sigma}$.
$T_{pd}$, $T'_{pd}$, and $T''_{pd}$ are the hopping integrals between $3d_{x^2-y^2}$ and $2p_\sigma$,  between $3d_{3z^2-r^2}$ and $2p_\sigma$, and between $3d_{3z^2-r^2}$ and $2p_z$, respectively, with $T_{pd}=\sqrt{3}T'_{pd}=\frac{\sqrt{3}}{2}T''_{pd}$.
The tetragonality of NiO$_6$ octahedron is represented by introducing $\alpha$, where $0<\alpha<1$ for elongated octahedron along the z direction.
The parameters $\varepsilon_d$ ($\varepsilon_p$) and $U_d$ are the energy level of Cu$3d_{x^2-y^2}$ (O$2p$) and the Coulomb repulsion of Cu$3d_{x^2-y^2}$ orbital, respectively.
$\varepsilon_\mathrm{Ni}$, $U_\mathrm{Ni}$, $U'_\mathrm{Ni}$, and $K_\mathrm{Ni}$ are the energy levels of Ni$3d$ orbitals, intra- and inter-orbital Coulomb repulsions, and the exchange interaction, respectively, with $U_\mathrm{Ni}=U'_\mathrm{Ni}+2K_\mathrm{Ni}$. Hereafter we set $\varepsilon_d=0$.

Following a procedure by Zhang and Rice~\cite{zhang}, let us introduce three kinds of O$2p$ Wannier orbitals, i.e., symmetric, antisymmetric, and nonbonding.
Among them, we can neglect the nonbonding orbital since it is decoupled from other orbitals.
For a system with $N$ unit cells, the symmetric and antisymmetric Wannier orbitals are given by
$
\phi^s_{\mathbf{i}\sigma}=-iN^{-\frac{1}{2}}\sum_{\mathbf{k}}e^{i\mathbf{k\cdot i}}\beta_{s\mathbf{k}}
(S_x p_{x\mathbf{k}\sigma}-S_y p_{y\mathbf{k}\sigma}),
$ and $
\phi^a_{\mathbf{i}\sigma}=iN^{-\frac{1}{2}}\sum_{\mathbf{k}}e^{i\mathbf{k\cdot i}}\beta_{a\mathbf{k}}
[\sqrt{2}S_x S_y(S_y p_{x\mathbf{k}\sigma}+S_x p_{y\mathbf{k}\sigma})-i\alpha(S_x^2 + S_y^2)p_{z\mathbf{k}\sigma}],
$
where $S_{x(y)}=\sin\frac{k_{x(y)}}{2}$, $\beta_{s\mathbf{k}}=(S_x^2+S_y^2)^{-\frac{1}{2}}$,
$\beta_{a\mathbf{k}}=\beta_{s\mathbf{k}}(2S_x^2S_y^2+\alpha^2\beta_{s\mathbf{k}}^{-2})^{-\frac{1}{2}}$, and the operators $p_{x\mathbf{k}\sigma}$, $p_{y\mathbf{k}\sigma}$, and $p_{z\mathbf{k}\sigma}$ are the Fourier transformations of $p_{\mathbf{i+\frac{x}{2}}\sigma}$, $p_{\mathbf{i+\frac{y}{2}}\sigma}$, and $\frac{1}{\sqrt{2}}(p_{\mathbf{i+z}\sigma}-p_{\mathbf{i-z}\sigma})$, respectively.
Using these Wannier orbitals, (1) reads
\begin{eqnarray}\label{dp3}
H_T&&=2T_{pd}\sum_{\mathbf{ij}\sigma}\tau_\mathbf{ij}d^\dag_{\mathbf{i}\sigma}\phi^s_{\mathbf{j}\sigma}
-2T'_{pd}\sum_{\mathbf{i_0j}\sigma}\tau'^s_\mathbf{i_0j}d'^\dag_{\mathbf{i_0}\sigma}\phi^s_{\mathbf{j}\sigma}
\nonumber\\&&
+2\sqrt{2}T'_{pd}\sum_{\mathbf{i_0j}\sigma}\tau'^a_\mathbf{i_0j}d'^\dag_{\mathbf{i_0}\sigma}\phi^a_{\mathbf{j}\sigma}
+\mathrm{H.c.} \nonumber\\&&
+\varepsilon_p\sum_{\mathbf{i}\sigma\eta=s,a}\phi^{\eta\dag}_{\mathbf{i}\sigma}\phi^\eta_{\mathbf{i}\sigma}
+\varepsilon_\mathrm{Ni}\sum_{\mathbf{i}_0\sigma\gamma=d,d'}n^\gamma_{\mathbf{i}_0\sigma}
,
\end{eqnarray}
where
$\tau_\mathbf{ij}=\frac{1}{N}\sum_\mathbf{k}\beta^{-1}_{s\mathbf{k}}e^{i\mathbf{k\cdot(i-j)}}$,
$\tau'^s_\mathbf{ij}=\frac{1}{N}\sum_\mathbf{k}\beta_{s\mathbf{k}}(S_x^2 -S_y^2)e^{i\mathbf{k\cdot(i-j)}}$, and
$\tau'^a_\mathbf{ij}=\frac{1}{N}\sum_\mathbf{k}\beta_{s\mathbf{k}}^2\beta_{a\mathbf{k}}^{-1}e^{i\mathbf{k\cdot(i-j)}}$.
Their value is the largest if $\mathbf{i}-\mathbf{j}$ is small: for example, $\tau_{\mathbf{i}=\mathbf{j}}=0.958$, $\tau'^s_{\mathbf{i}=\mathbf{j}}=0$, $\tau'^s_{\mathbf{i}=\mathbf{j+x}}=-0.258$, and $\tau'^a_{\mathbf{i}=\mathbf{j}}=0.915$.
In the case of a single Ni impurity in the system, it is convenient to introduce a new antisymmetric operator given by $\tilde{\phi}_{a\sigma} = \tilde{\beta}^{-1} \sum_{\mathbf{j}} \tau'^a_\mathbf{i_0j} \phi^a_{\mathbf{j}\sigma}$ with $\tilde{\beta}^2 = \sum_{\mathbf{j}}|\tau'^a_\mathbf{i_0j}|^2$, since $\phi^a_{\mathbf{j}\sigma}$ only couples to the impurity.
Then, the number of the orbitals in the system becomes $2N+2$.

We use the Lanczos-type exact diagonalization technique on a $N=\sqrt8\times\sqrt8$-unit-cell cluster with a single Ni impurity under periodic boundary conditions~\cite{note1}.
The parameter values have been estimated to be  $T_{pd}\simeq0.95\sim 1.3$~eV, $U_d\simeq8\sim 10.5$~eV, $\varepsilon_p \simeq2\sim3.5$~eV~\cite{maekawa}.
In the present study, we take $T_{pd} = 1$~eV, $U_d = U_\mathrm{Ni} = 8$~eV, $K_\mathrm{Ni} = 0.8$~eV, and $\varepsilon_p = 3$ eV.
$\varepsilon_\mathrm{Ni}$ is an unknown parameter and $\alpha$ is taken to be $1/\sqrt{2}$ for simplicity.

Insulating cuprates are known to be charge-transfer (CT) type, where the CT energy $\Delta$ between Cu and O is smaller than $U_d$~\cite{zaanen}.
Nickel oxides also belong to the CT type, but the charge transfer energy $\Delta_\mathrm{Ni}$ is larger than $\Delta$~\cite{bocquet}. 
$\Delta$ is given by $\Delta = E (d^{10}\underline{L})-E (d^9)=\varepsilon_p$, where $E$ is the energy of a Cu-O unit for a given configuration without hopping terms.
Similarly $\Delta_\mathrm{Ni} = E(d^9\underline{L})-E(d^8) = \Delta -\varepsilon_\mathrm{Ni} -U_\mathrm{Ni} +3K_\mathrm{Ni}$.
We assume that $\Delta_\mathrm{Ni}$ for the Ni impurity is similar to the bulk systems, since the CT energy is predominantly determined by the energy level of $3d$ orbitals when ligand and environment are the same~\cite{bocquet}.
Therefore, we take $\Delta_\mathrm{Ni}>\Delta$ in the present study of Ni impurity, leading to $\varepsilon_\mathrm{Ni}<-5.6$~eV.

\begin{figure}[t]
\includegraphics[width=7.5cm]{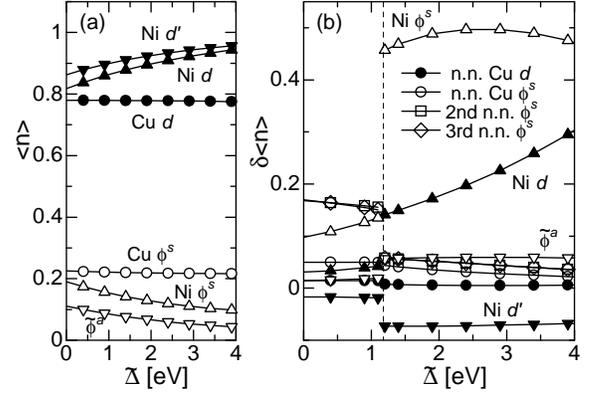}
\caption{\label{al07h0diff}
(a) Averaged hole number $\langle n \rangle$ on each orbital in undoped ground state of an 8-unit-cell cluster as a function of the CT energy difference $\tilde{\Delta}\equiv\Delta_\mathrm{Ni}-\Delta$. 
Filled upward and downward triangles represent Ni~$3d_{x^2-y^2}$ (denoted as $d$) and $3d_{3z^2-r^2}$ ($d'$), respectively. $\phi^s$ and $\tilde{\phi}^a$ on Ni site are shown by open triangles. Filled (open) circles denote Cu~$d$ (Cu~$\phi^s$) orbitals at the nearest neighbor (n.n.) sites from Ni.
 (b) The difference of $\langle n \rangle$ between one-hole-doped and undoped ground states.
Open squares and diamonds denote $\phi^s$ orbitals at the second and third n.n. sites, respectively.
Other symbols are the same as (a). 
The vertical broken line represents the position where the ground-state symmetry changes.
}
\end{figure}

Figure~\ref{al07h0diff}(a) shows hole number $\langle n \rangle$ on each orbital of the cluster as a function of the difference of the CT energies $\tilde{\Delta}\equiv\Delta_\mathrm{Ni}-\Delta$ in the undoped case (9 holes in the 8-unit-cell system).
We find that the total number of holes at the Ni site (the sum of Ni~$d$, Ni~$d'$, Ni~$\phi^s$ and $\tilde{\phi}^a$) is almost two (1.99 at $\tilde{\Delta}=0.4$~eV and 2.04 at $\tilde{\Delta}=3.9$~eV), which means that nominally Ni$^{2+}$ is formed. The local spin state is of high spin ($S=1$). Remaining holes are distributed on Cu~$d$, Cu~$\phi^s$ orbitals, forming Cu$^{2+}$ on each Cu site.
Note that $\langle n \rangle$ on the Cu orbitals at the second and third nearest-neighbor (n.n.) sites from Ni is nearly the same as those of the n.n. sites. 

Next we examine how a hole introduced into the system is distributed among the orbitals.
In the range of $\tilde{\Delta}$ shown in Fig.~\ref{al07h0diff}, we find that with increasing $\tilde{\Delta}$ the ground state with one additional hole (10 holes in total) changes quantum number at $\tilde{\Delta}\sim1.2$~eV from the total spin $S=1$ and A$_2$ irreducible representation of $C_{4v}$ (degenerated with B$_1$) to $S=0$ and B$_1$.
The difference of hole number, $\delta\langle n \rangle$, on each orbital between the one-hole doped and undoped ground states is plotted in Fig.~\ref{al07h0diff}(b). 
We find that for $\tilde{\Delta} \gtrsim 1.2$ a hole introduced in the system occupies predominantly on Ni~$\phi^s$ orbital, i.e., an oxygen Wannier orbital that mainly couples to Ni$3d_{x^2-y^2}$. In total, more than 60\% of the hole enters into Ni-related orbitals. This is consistent with experimental suggestions~\cite{bhat,nakano,hiraka2,matsuda,hiraka3,hiraka} that a Ni impurity may bind a hole and form the ZR doublet.
Note that negative $\delta\langle n \rangle$ for Ni~$d'$ is the consequence of strong energy gain by accommodating holes into the orbitals with $x^2-y^2$ symmetry.
For $\tilde{\Delta}\gtrsim 1.2$, the doped hole is distributed mainly on oxygen orbitals away from the Ni impurity.
Since an effective CT energy for Ni oxides has been reported to be $\sim5$~eV~\cite{bocquet}, it is reasonable to consider that $\tilde{\Delta}>1.2$~eV.
Therefore, the present result strongly supports the binding of a hole around a Ni impurity in real Ni-substituted cuprates.

In order to confirm the stability of the ZR doublet, we compared binding energy of the ZR doublet on a NiO$_6$ cluster with that of the ZR singlet on a CuO$_4$ cluster.
We found that the ZR doublet is more stable than the ZR singlet when $\Delta_\mathrm{Ni}$ is larger than $\Delta$ by 0.3~eV.
This is qualitatively consistent with the data shown in Fig.~\ref{al07h0diff}(b), implying that the stability of the ZR doublet is governed by local character around Ni.
In the limit of ($U_d$, $U_\mathrm{Ni}$, $\Delta$, $\Delta_\mathrm{Ni}$) $\gg T_{pd}$ that is approximately satisfied in the present parameter set, the binding-energy difference of the doublet and the singlet, $\delta E_\mathrm{B}$, may be given by~\cite{note2}
\begin{equation}
\delta E_\mathrm{B}=2T_{pd}^2\left(\frac{4}{U_d-\Delta}+\frac{2}{\Delta}-\frac{3}{U_\mathrm{Ni}+K_\mathrm{Ni}-\Delta_\mathrm{Ni}}-\frac{1}{\Delta_\mathrm{Ni}}\right).
\label{Binding}
\end{equation}
We notice that dominant negative contribution to $\delta E_\mathrm{B}$ comes from the third term in (\ref{Binding}).
This implies the importance of the condition that $U_\mathrm{Ni}$ ($\sim$8~eV) is not much larger than $\Delta_\mathrm{Ni}$ ($\sim$5~eV).

\begin{figure}[t]
\includegraphics[width=6cm]{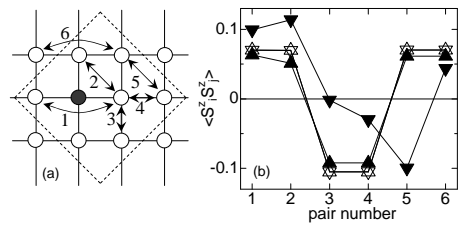}
\caption{\label{corr}
(a) Schematic diagram for an 8-unit-cell cluster (the inside of dotted lines) where the filled and open circles denote Ni and Cu sites.
(b) Spin-spin correlation function for the pairs numbered in (a).
The upward (downward) triangles show data for $\tilde{\Delta}=1.4$ $(0.9)$~eV, and the open and filled symbols denote the undoped and one-hole-doped cases, respectively.
}
\end{figure}

When a hole is bound to a Ni impurity, localized Cu spins are expected to be unaffected and thus AF correlation may remain. 
Defining $S^z_\mathbf{i} = ( n^d_{\mathbf{i},\uparrow} - n^d_{\mathbf{i},\downarrow} + \phi^{s\dag}_{\mathbf{i},\uparrow}\phi^{s}_{\mathbf{i},\uparrow} - \phi^{s\dag}_{\mathbf{i},\downarrow}\phi^{s}_{\mathbf{i},\downarrow})/2$, we show in Fig.~\ref{corr} the equal-time spin-spin correlation function $\langle S^z_\mathbf{i} S^z_\mathbf{j} \rangle$ between two Cu sites in undoped and one-hole doped cases with $\tilde{\Delta} = 1.4$ and $0.9$ eV.
Here, $\tilde{\Delta}$ is chosen near the boundary of two regions shown in Fig.~\ref{al07h0diff}(b).
In the undoped case, the correlation on the same (different) sub-lattice is positive (negative), implying the presence of AF order. 
For $\tilde{\Delta}=1.4$~eV, $\langle S^z_\mathbf{i} S^z_\mathbf{j} \rangle$ in the one-hole doped case is very similar to the undoped case.
This is a natural consequence of hole binding around Ni impurity.
On the other hand, the correlation is dramatically changed from the undoped case when $\tilde{\Delta}=0.9$~eV. This is due to the destruction of AF order caused by carrier motion.

To directly observe hole binding around Ni impurity, we need to use site-selective probes.
One of them would be XAFS that has been performed recently~\cite{hiraka}.
As another probe, we propose a RIXS experiment for Ni $K$-edge, which can detect charge excitations related to the bound hole.
In the Ni $K$-edge RIXS, the emission of a photon with a dipole transition between Ni~$4p$ and Ni~$1s$ states occurs resonantly by tuning the energy of incoming photon to Ni $K$ absorption edge.
In the intermediate state, we introduce Coulomb interaction between $3d$ and $1s$-core holes~\cite{tsutsui}, given by
$H_{1s\mathrm{-}3d} = U_c \sum_{\mathbf{i}_0,\gamma,\sigma,\sigma'} n_{\mathbf{i}_0\sigma}^\gamma n_{\mathbf{i}_0\sigma'}^s$, where $n_{\mathbf{i}_0\sigma}^s$ is the number operator of $1s$-core hole.
By assuming that the $4p$ photo-electron enters into the bottom of the $4p$ bands~\cite{tsutsui}, the RIXS spectrum for Ni $K$-edge as well as Cu $K$-edge is expressed as
$
I(\Delta\omega)=\sum_{f}\left|\langle f| D_{\mathbf{K}_f}^\dag G(\omega_i)
D_{\mathbf{K}_i} |0\rangle\right|^2
\delta\left(\Delta\omega-E_f+E_0\right),
$
where $D_\mathbf{K}=\sum'_{\mathbf{i},\sigma} e^{i\mathbf{K}\cdot \mathbf{i}} p_{4p,\mathbf{i}\sigma}^\dag s_{\mathbf{i}\sigma}^\dag + {\rm H.c.}$ with the creation operator $s^\dagger_{\mathbf{i}\sigma}$ ($p^\dagger_{4p,\mathbf{i}\sigma}$) of $1s$-core hole ($4p$ electron).
$\sum'_{\mathbf{i}}$ denotes summation over Ni (Cu) sites for Ni (Cu) $K$-edge. 
$\mathbf{K}_{i(f)}$ is the wave vector of the incoming (outgoing) photon with energy $\omega_{i(f)}$, and $\Delta\omega=\omega_i-\omega_f$.
$G^{-1}(\omega_i)=\omega_i+i\Gamma-H_{dp}-H_{1s{\rm -}3d}-H_{1s,4p}$, where $H_{1s,4p}$ is composed of the energy separation $\varepsilon_{1s{\rm -}4p}^\mathrm{Ni(Cu)}$ between the Ni (Cu) $1s$ level and the bottom of the $4p$ band, and $\Gamma$ is the inverse of relaxation time in the intermediate state.
$|0\rangle$ is the ground state with energy $E_0$, $|f\rangle$ is the final state of RIXS with energy $E_f$.
We use $\Gamma=1$~eV and $U_c=4$~eV~\cite{ide}.
$I(\Delta\omega)$ is calculated by using a modified version of the conjugate-gradient method together with the Lanczos technique.

\begin{figure}[t]
\includegraphics[width=6cm]{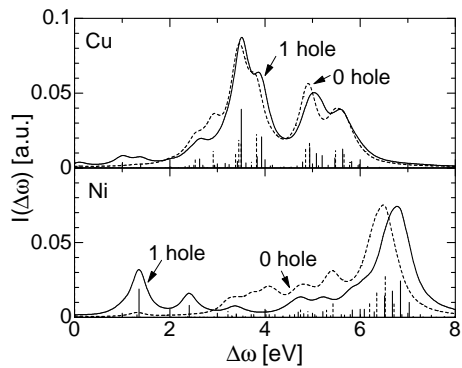}
\caption{\label{rixsfig}
RIXS spectra for Cu (upper panel) and Ni (lower panel) $K$-edges in an 8-unit-cell cluster with a Ni impurity.
The solid and broken lines denote the undoped and one-hole-doped cases, respectively.
$\tilde{\Delta} = 1.4$~eV.
The $\delta$-functions (the vertical lines) are convoluted with a Lorentzian broadening of $0.2$~eV.
}
\end{figure}

Figure~\ref{rixsfig} shows Cu $K$-edge (upper panel) and Ni $K$-edge (lower panel) RIXS spectra in undoped and one-hole-doped cases for $\tilde{\Delta}=1.4$~eV.
The spectra at zero momentum transfer ($\mathbf{K}_f-\mathbf{K}_i=0$) are shown for Cu $K$-edge.
In the undoped case, the edge of the CT gap is located at $\sim2.1$ eV and $\sim3.2$ eV for Cu and Ni, respectively.
The difference comes from different CT energies.
This is qualitatively consistent with experimental observations for La$_2$CuO$_4$ ($\sim2.2$ eV) and La$_2$NiO$_4$ ($\sim4$ eV)~\cite{collart}.
Furthermore, recent RIXS experiments for Ni-substituted La$_2$CuO$_4$ also show similar behaviors~\cite{ishiiimp}.
In Cu $K$-edge RIXS, the spectrum around 4.5-6~eV is associated with the excitations from bonding to antibonding states in CuO$_4$ plaquette~\cite{hill,ide}.
A similar structure appears in the energy region from 6 to 7~eV for Ni $K$-edge RIXS.

The Ni $K$-edge spectrum is strongly affected by hole-doping as shown in the lower panel of Fig.~\ref{rixsfig}.
In particular, new spectral structures appear at $\Delta\omega\sim 1.35$~eV and $\sim2.13$~eV within the CT gap.
Examining eigenstates generating the new structures, we find that the hole number of Cu~$\phi^s$ is larger than that in the ground state, while the hole number at the Ni site is smaller.
This means that the ZR singlet on Cu are dominating in the corresponding excited states.
Since the hole number at the Ni site is large in the ground state due to the formation of the ZR doublet, the new spectral structures are associated with excitations from the ZR doublet to ZR singlet.
Therefore, if we observe spectral weights inside the CT gap for Ni-substituted cuprates, it can be identified as direct evidence of hole bounding around Ni impurity. Note that the spectrum for Cu $K$-edge is almost unchanged upon hole-doping although a small hump is seen around 1~eV.

We have neglected direct O-O hoppings in (\ref{dp}).
The hoppings may cause the decrease of $\Delta$, leading to less stability of the ZR doublet as seen from (\ref{Binding}).
However, we found that the boundary at $\tilde\Delta\sim0.3$~eV for single clusters without the hopping parameter $T_{pp}$ increases only by 0.1~eV for a realistic value of $T_{pp}$ (=0.5~eV).

The formation of the ZR doublet at Ni site imposes restrictions on modeling impurity effects in cuprate superconductors, in particular, on the construction of single-band models like $t$-$J$-type model~\cite{poilblanc,ohta,riera,kuroda,tsuchiura,xiang}.
We should treat a Ni impurity as a $S=\frac{1}{2}$ spin in hole-doped cases.
The spin is expected to couple to the neighboring Cu spins antiferromagnetically with an interaction parameter determined by competition between superexchange processes and the process of the motion of the $\phi^s$ hole.
Other holes that hop to the impurity position may feel repulsive interaction.
This model is close the those used in Ref.~\cite{poilblanc}.
More precise effective $t$-$J$-type model is now under construction.

In summary, we have carried out the numerically exact diagonalization on the eight-unit-cell cluster with a Ni impurity site representing Ni-substituted cuprate superconductors.
Using realistic parameters of the cuprates, we have found that a hole can be bound in the NiO$_4$ plaquettes forming the ZR doublet.
This is due to the fact that the ZR doublet is more stable in energy than the ZR singlet on Cu.
This finding supports theoretically recent experimental suggestion of the presence of the ZR doublet.
Also this imposes strong restrictions on modeling Ni-substituted cuprates.
We have proposed that the hole binding can be seen in RIXS for Ni $K$-edge.
We hope that the proposed RIXS experiment will be done in the near future.

We are grateful to H. Hiraka, K. Ishii, J. Mizuki, K. Yamada, and W. Koshibae for helpful discussions.
This work was supported by Nanoscience Program of Next Generation Supercomputing Project and Grant-in-Aid for Scientific Research from MEXT, and the inter-university cooperative research program of IMR, Tohoku Univ.
Computations were carried out in IMR, Tohoku Univ.; ISSP, Univ. of
Tokyo; and YITP, Kyoto Univ.


\begin{thebibliography}{00}
\bibitem{note1}We also examined $N=2$ and $N=2\times2$ clusters and found that size dependence is small. The $N=8$ cluster is shown in Fig.~\ref{corr}(a).
\bibitem{note2} $\Delta_{\rm Ni}$ of $\delta E_{\rm B}=0$ in (\ref{Binding}) is $\sim 6.5$~eV, which is much larger than that estimated exactly from the single cluster.
\end{thebibliography}

\end{document}